\documentclass[twocolumn,letter]{jpsj3}
\usepackage{graphicx}
\usepackage{txfonts,bm,revsymb}
\usepackage{bm}
\usepackage{color}

\def\Vec#1{\bm{#1}}

\title{
Surface states around a vortex in topological superconductors: Intersection of a surface and a vortex}

\author{Yuki Nagai, Hiroki Nakamura, and Masahiko Machida}
\inst{
\address{CCSE, Japan  Atomic Energy Agency, 178-4-4, Wakashiba, Kashiwa, Chiba 277-0871, Japan} 
}

\abst{
We numerically show that 
the zero-energy Majorana surface states are suppressed around a vortex  in the three-dimensional topological superconductors such as Cu$_{x}$Bi$_{2}$Se$_{3}$ and Sn$_{1-x}$In$_{x}$Te. 
On the other hand, the zero-energy Majorana bound states along the vortex line are robust against cut by the surface. 
The suppression of the surface bound states is similar to that with a magnetic impurity on the surface 
of the topological insulator.
The suppression of the surface-bound states around a vortex can be observed as the unconventional energy dependence of imaging of the scanning tunneling microscopy/spectroscopy. 
}

\begin{document}
\maketitle

Majorana fermions predicted an over seventies years ago by E. Majorana\cite{Majorana} have attracted much attention in both high-energy and condensed matter physics.
A neutrino is a candidate of Majorana fermion though no direct evidence is found in high-energy physics. 
The discovery of a topological superconductor opened a new research avenue to investigate the Majorana fermion in materials.\cite{Kane,HK_RMP10,Bernevig,Fu,Moore, Konig,FuKane,Nishide,TSato,Kuroda,Shen2,Nagaivortex}
This zero-energy Majorana fermion has a counterintuitive property that a particle and anti-particle are identical. 
Gapless zero-energy quasi-particles at surfaces in superconductors are regarded as the Majorana fermions. 
The superconducting topological insulators Cu$_{x}$Bi$_{2}$Se$_{3}$ and Sn$_{1-x}$In$_{x}$Te present a chance to investigate {\it bulk} topological superconductivity in real bulk materials\cite{Hor_L10,Wray_NP10,MKR_L11,FB_L10,Sasaki,Kirzhner,SasakiSn}, not in heterostructures.

In the case of the topological superconductor, the spatial dimension of Majorana fermions $M$ is less than the dimension of a system $N$ ($M < N$), since Majorana states 
always appear on a boundary. 
The boundary Majorana fermion appears at zero-energy of gapless linear dispersion (Dirac cone).
For example, in three-dimensional systems, a surface Majorana fermion and a vortex Majorana fermion 
appear at zero-energy of a two- and one- dimensional Dirac cone, respectively.
Each gapless states are {\it topologically} protected. 
In the type DIII three-dimensional topological superconductor\cite{Schnyder,Nishida}, which is the candidate of the superconducting pairing 
in Cu$_{x}$Bi$_{2}$Se$_{3}$ and Sn$_{1-x}$In$_{x}$Te, the time-reversal symmetry protects the surface states 
topologically. 
The zero-energy bound states also appear around a vortex, since the vortex is regarded as the topological defect\cite{Teo}. 
The three-dimensional system can form an intersection point of a surface and a vortex.  
Therefore, the questions arise: What is happened at the intersection point? Which are Majorana fermions stronger?
These questions are important in terms of the robustness of the Majorana fermions in topological quantum computing utilizing non-Abelian anyon statistics\cite{TeoPRL}.

In the two-dimensional $p$-wave triplet superconductor, which is a topological superconductor, 
the emergence of the Majorana bound states inside 
the vortex core has been numerically confirmed by using Bogoliubov-de Gennes (BdG) formalism\cite{Takigawa,Matsumoto}.
However, in a two-dimensional system, we cannot construct the intersection between a surface and a vortex line perpendicular to the surface.
Therefore, in order to investigate the robustness of the surface- and vortex- bound states, 
it is important to consider the three-dimensional bulk superconductor.
In addition, recently,  the scanning tunneling microscopy/spectroscopy (STM/STS) was applied to Cu$_{x}$Bi$_{2}$Se$_{3}$\cite{Levy}. 
The behavior of quasiparticles at the intersection can be detected experimentally.

In this paper, we numerically reveal that the surface-bound states are suppressed 
around a vortex, which is the intersection point of two different Majorana bound states. 
We also show that the gap-less vortex Majorana bound states are  robust against the surface. 
The suppression of the surface bound states is similar to that with a magnetic impurity on the surface 
of the topological insulator.
The suppression of the surface-bound states around a vortex can be observed as the unconventional energy dependence of imaging of the scanning tunneling microscopy/spectroscopy.

\begin{figure}[ht]
  \begin{center}
    \begin{tabular}{p{0.5 \columnwidth}p{0.5 \columnwidth}}
      \resizebox{0.5\columnwidth}{!}{\includegraphics{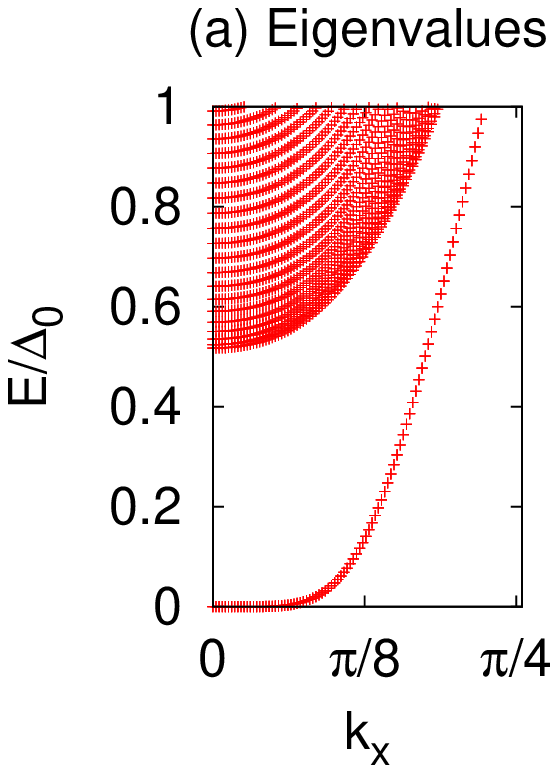}} &
      \resizebox{0.5\columnwidth}{!}{\includegraphics{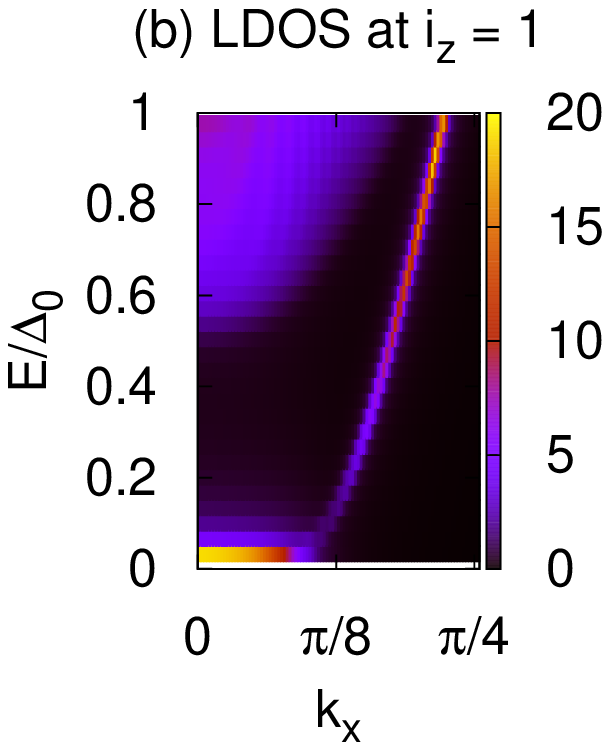}} \\
    \end{tabular}
\caption{\label{fig:1d}
(Color online)(a)Eigenvalue distribution and (b) the LDOS on the top surface in the one-dimensional tight binding model without a vortex. 
}
  \end{center}
\end{figure}

\begin{figure}[thb]
\begin{center}
\includegraphics[width = 0.8\columnwidth]{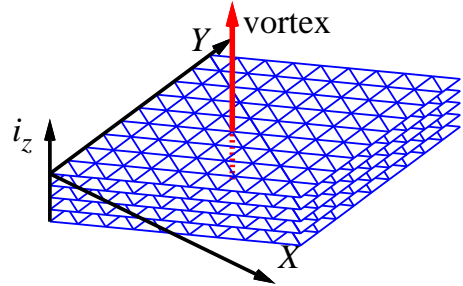}
\caption{\label{fig:figm}
Schematic figure of the three-dimensional tight-binding lattices. The single vortex is parallel to the $z$-axis. 
}
\end{center}
\end{figure}

We start with a model Hamiltonian of the topological insulator Bi$_{2}$Se$_{3}$ proposed by several groups. 
We note that this model in the normal states is described by the three-dimensional Dirac Hamiltonian.
Our result can be applied to the case of SnTe, since this material can be modeled by the three-dimensional Dirac Hamiltonian shifting origin in momentum space to $L$-point\cite{Hsieh}. 
In order to examine the surface-bound states and vortex-bound states in the model, we 
have the three-dimensional mean-field Hamiltonian on $L_{x} \times L_{y}$ triangle lattice and discrete $L_{z}$ planes 
based on the Bogoliubov-de Gennes (BdG) formalism,
\begin{align}
H =  \sum_{i,j} 
\left(\begin{array}{cc}c_{i}^{\dagger} & c_{i}^{T}
\end{array}\right)
\left(\begin{array}{cc}\hat{H}_{ij} & \hat{\Delta} f({\bm R}_{i}) \delta_{ij} \\
\hat{\Delta}^{\dagger} \delta_{ij} f({\bm R}_{i})^{\ast}& - \hat{H}_{ij}^{\ast}
\end{array}\right)
\left(\begin{array}{c}c_{j}  \\c_{j}^{\ast}
\end{array}\right),
\end{align}
where $c_{i}$ is the four-component annihilation operator at the lattice site $i = (i_{x},i_{y},i_{z})$. 
${\bm R}_{i}$ is defined by ${\bm R}_{i} \equiv (X,Y)$ with $X = (\sqrt{3}/2) (i_{x} - L_{x}/2)$ and $Y = (i_{x}-L_{x}/2)/2+(i_{y} -L_{y}/2)$. The normal-states Hamiltonian $\hat{H}_{ij}$ in real space is given by Fourier-transforming the Hamiltonian in momentum space expressed as 
\begin{align}
\hat{H}(\Vec{k}) &\equiv M(\Vec{k}) \gamma^{0} + \sum_{i=0}^{3} P_{i}(\Vec{k}) \gamma^{0} \gamma^{i},
\end{align}
where, 
\begin{align}
M({\bm k}) &\equiv M_{0} - 2 \bar{B}_{1}(1 - \cos (k_{z}))- \bar{B}_{2}\eta({\bm k}_{\perp}), 
\end{align}
\begin{align}
P_{0}({\bm k}) &\equiv 2 \bar{D}_{1}(1 - \cos (k_{z})) + \bar{D}_{2} \eta({\bm k}_{\perp}) - \mu, \\
P_{1}({\bm k}) &\equiv \frac{2}{3} \bar{A}_{2} \sqrt{3} \sin \left( \frac{\sqrt{3}}{2} k_{x} \right) \cos \left(\frac{k_{y}}{2} \right), \\
P_{2}({\bm k}) &\equiv \frac{2}{3}  \bar{A}_{2} \left(\cos \left( \frac{\sqrt{3}}{2} k_{x} \right) \sin \left(\frac{k_{y}}{2} \right) + \sin(k_{y})\right), \\
P_{3}(k_{z}) &\equiv \bar{A}_{1} \sin(k_{z}), 
\end{align}
with 
$\eta({\bm k}) \equiv (3 - 2 \cos(\sqrt{3} k_{x}/2)\cos(k_{y}/2) - \cos(k_{y}))$.
The $\gamma^{i}$ denotes the $4 \times 4$ gamma matrices in the Dirac representation, which can 
be described as $\gamma^{0} = \hat{\sigma}_{z} \otimes 1$ and  $\gamma^{i = 1,2,3} = i \hat{\sigma}_{y} \otimes \hat{s}_{i}$ with $2 \times 2$ Pauli matrices $\hat{\sigma}_{i}$ in the orbital space and $\hat{s}_{i}$ in the spin space. 
$\hat{\Delta}$ is a $4 \times 4$ matrix whose elements are given as $\Delta_{\sigma \sigma'}^{l m}$ using the 
orbital $l(m)$and spin $\sigma(\sigma')$ indices. 
We consider the fully-gapped topological gap function  (so-called $\hat{\Delta}_{2}$, inter-orbital spin-singlet gap function shown as\cite{Sasaki,NagaiMajo} 
\(
\Delta^{12}_{\uparrow\downarrow}
=
- \Delta^{12}_{\downarrow\uparrow}
= 
\Delta^{21}_{\uparrow\downarrow}
=
- \Delta^{21}_{\downarrow\uparrow}
=1
\), 
 in Ref.~\citen{Sasaki}).
We consider the gap-amplitude $\Delta_{0} = 0.3$eV, and the chemical potential $\mu = 0.8$ eV.
Other parameters are $\bar{A}_{1}=1$ eV \AA, $\bar{A}_{2} = 1.5$ eV \AA, $M_{0} = -0.7$ eV, $\bar{B}_{1} = -0.5$ eV \AA$^{2}$, $\bar{B}_{2} = -0.75$ eV \AA$^{2}$, and $\bar{D}_{1}=\bar{D}_{2} = 0$ eV \AA$^{2}$, which are the same as those in Fig.~S10 in the supplemental materials of Ref.~\citen{Sasaki}. 
We show that the system with this parameter set has the zero-energy surface states and their intensity of the local density of states (LDOS) is strong as shown in Fig.~\ref{fig:1d}, with the use of the one-dimensional tight-binding model with $L_{z} = 64$ planes with the periodic boundary condition in $x$- and $y$- directions without a vortex ($f(\Vec{R}_{i}) = \Delta_{0}$)\cite{NagaiThermal}.
The Majorana surface bound states appear at the surface in this system. 
In this paper, we consider the three-dimensional tight-binding model with the vortex. 
The single vortex is located at $(i_{x},i_{y}) = (L_{x}/2,L_{y}/2)$ as shown in Fig.~\ref{fig:figm}. 
For simplicity, we do not solve the gap-equation but use a spatial distribution form of the order parameter around a single vortex $f({\bm R}_{i})$
 written as 
\begin{align}
f({\bm R}_{i}) &= e^{i \theta} \Delta_{0} \frac{|{\bm R}_{i}|}{\sqrt{|{\bm R}_{i}|^{2} + 1}},
\end{align}
where $\theta$ denotes the polar angle around $c$-axis, $\Delta_{0}$ is the amplitude of the order-parameter.
In order to consider both the surface and the vortex, 
one has to calculate an extreme large scale system, since the coherence length is long because of the small 
Fermi surface and the small superconducting gap. 
We consider $L_{x} = L_{y} = 96$ and $L_{z} = 32$.
This matrix dimension size is $96 \times 96 \times 32 \times 4 \times 2 = 2359296$.
To obtain the LDOS $n(E,i_{x},i_{y},i_{z})$ in this large space, we use 
the spectral polynomial expansion scheme\cite{NagaiJPSJ,Nagaivortex,Covaci} with the thousands of CPU cores.
We use renormalization factors $a = 30eV$, $b = -\mu$, a smearing factor $\eta = 1 \times 10^{-3}$eV and a cut-off 
parameter $n_{c}= 8000$, which are parameters described in Ref.~\citen{NagaiJPSJ}. 

\begin{figure}[t]
  \begin{center}
    \begin{tabular}{p{0.75 \columnwidth}}
      \resizebox{0.75 \columnwidth}{!}{\includegraphics{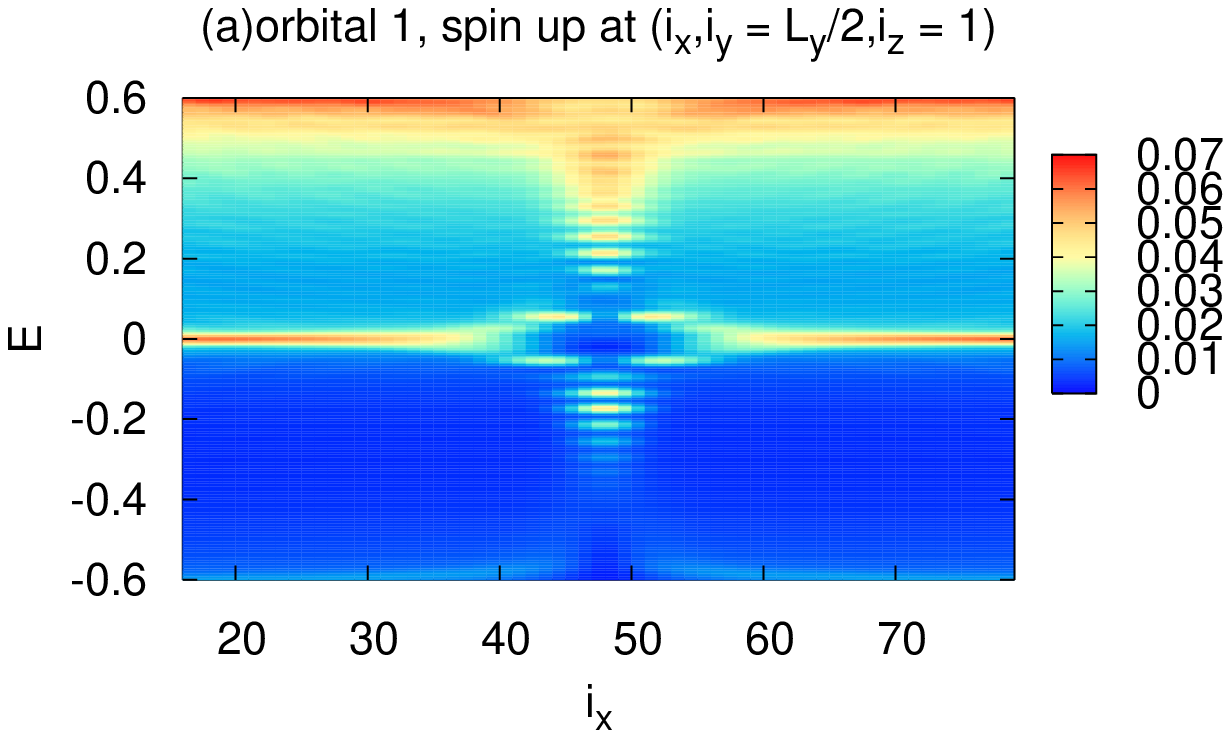}} \\
      \resizebox{0.75 \columnwidth}{!}{\includegraphics{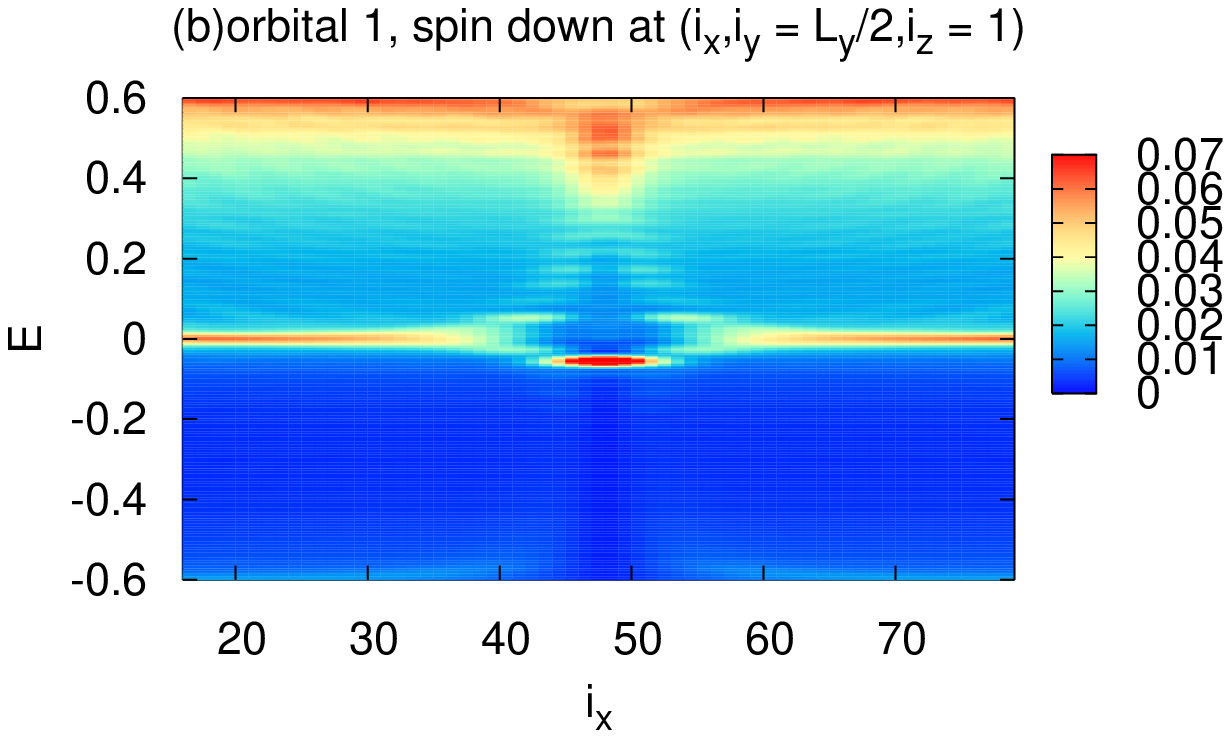}} \\
            \resizebox{0.75 \columnwidth}{!}{\includegraphics{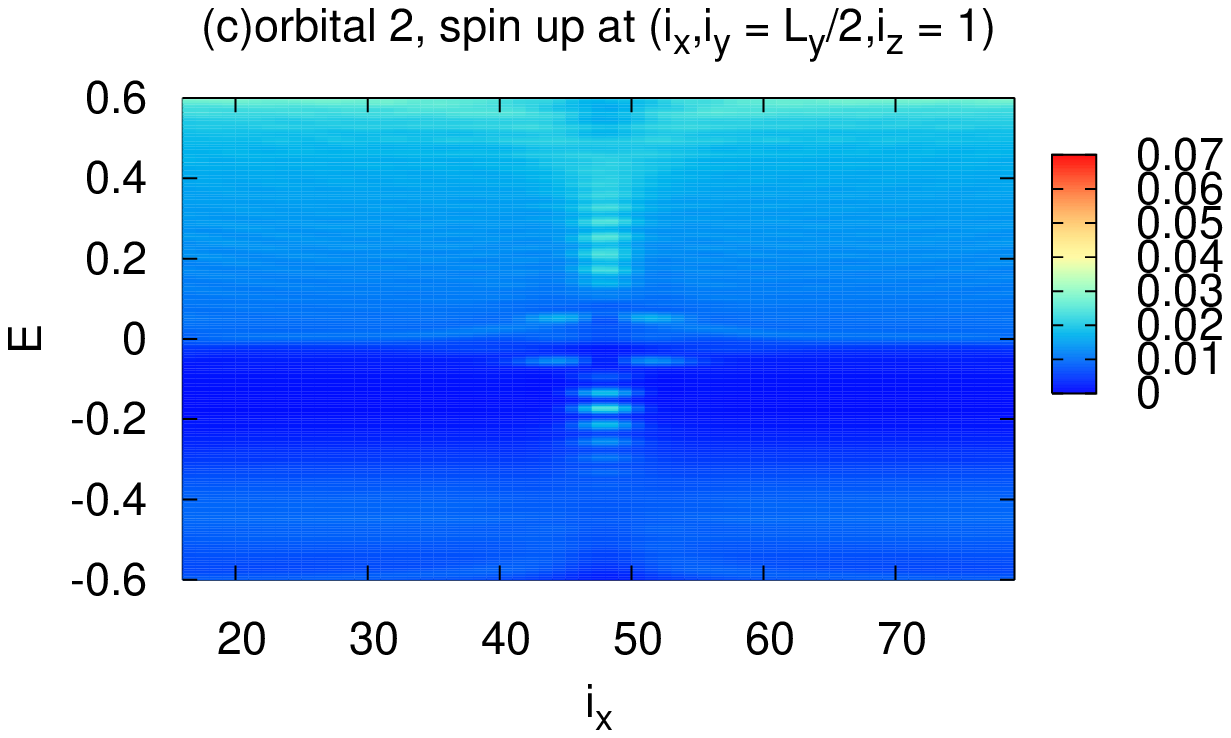}} \\
      \resizebox{0.75 \columnwidth}{!}{\includegraphics{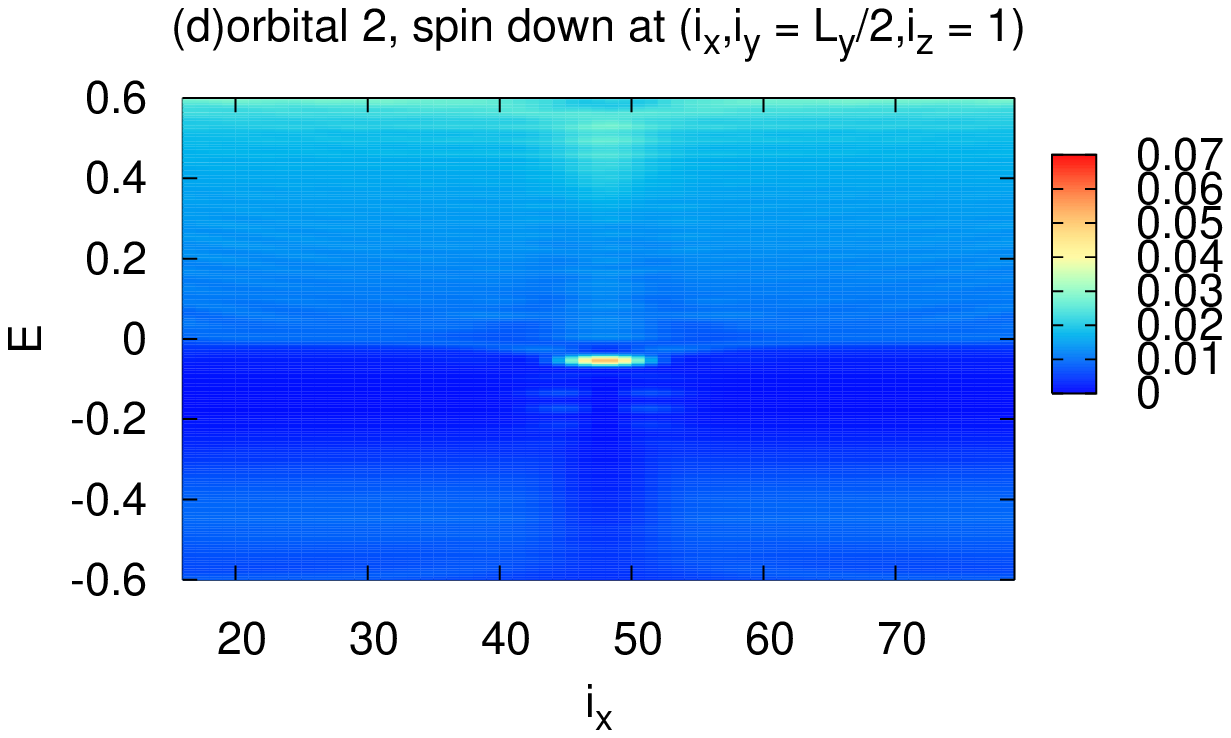}} \\
    \end{tabular}
\caption{\label{fig:fig3}
Energy dependence of the spin- and orbital- resolved local density of states around a vortex at $(i_{x},i_{y},i_{z}) = (i_{x},L_{y}/2,1)$. 
}
  \end{center}
\end{figure}

First, we focus on the surface bound states on the top of the three-dimensional lattice. 
There are two surfaces located at $i_{z} = 1$ and $i_{z} = L_{z}$. 
We calculate the energy dependence of the spin- and orbital- resolved LDOS at the surface $i_{z} = 1$ on the line $i_{y} = L_{y}/2$ as shown in Fig. ~\ref{fig:fig3}. 
We note that in the point-contact spectroscopy experiment observed the strong suppression of the zero-bias conductance peak (i.e. the zero energy LDOS) with a modest magnetic field\cite{Sasaki}.
They indicated that the helical Majorana fermions are naturally suppressed as the time-reversal symmetry is broken with 
the magnetic fields. 
Although the time-reversal symmetry is broken in our system, the zero-energy bound states appear far from a vortex.  
These zero-energy surface-bound states are orbital-polarized in this orbital basis. 
This behavior is consistent with the result shown in Fig.~\ref{fig:1d} in the one-dimensional tight-binding system without a vortex. 
Near the vortex core ($i_{x} = L_{x}/2$), the vortex opens up a local gap and suppresses the LDOS. 
The suppression of the surface bound states is similar to that with a magnetic impurity on 
the surface of the topological insulator\cite{Liu}. 
The vortex as the defect on the surface affects the surface Majorana fermions.
In the other words, the quasiparticles described by the massless Dirac dispersion on the surface locally obtain the mass around a magnetic object. 
Thus, the surface Majorana bound states are not robust against a vortex. 
We note that there is the finite-energy down-spin polarized bound state around the vortex on the surface at $E \sim -0.2 \Delta_{0}$.  
We consider that this finite-energy bound state have no relation with the zero-energy surface bound state since they have spin polarization opposite to each other.

\begin{figure}[t]
  \begin{center}
    \begin{tabular}{p{\columnwidth}}
      \resizebox{ \columnwidth}{!}{\includegraphics{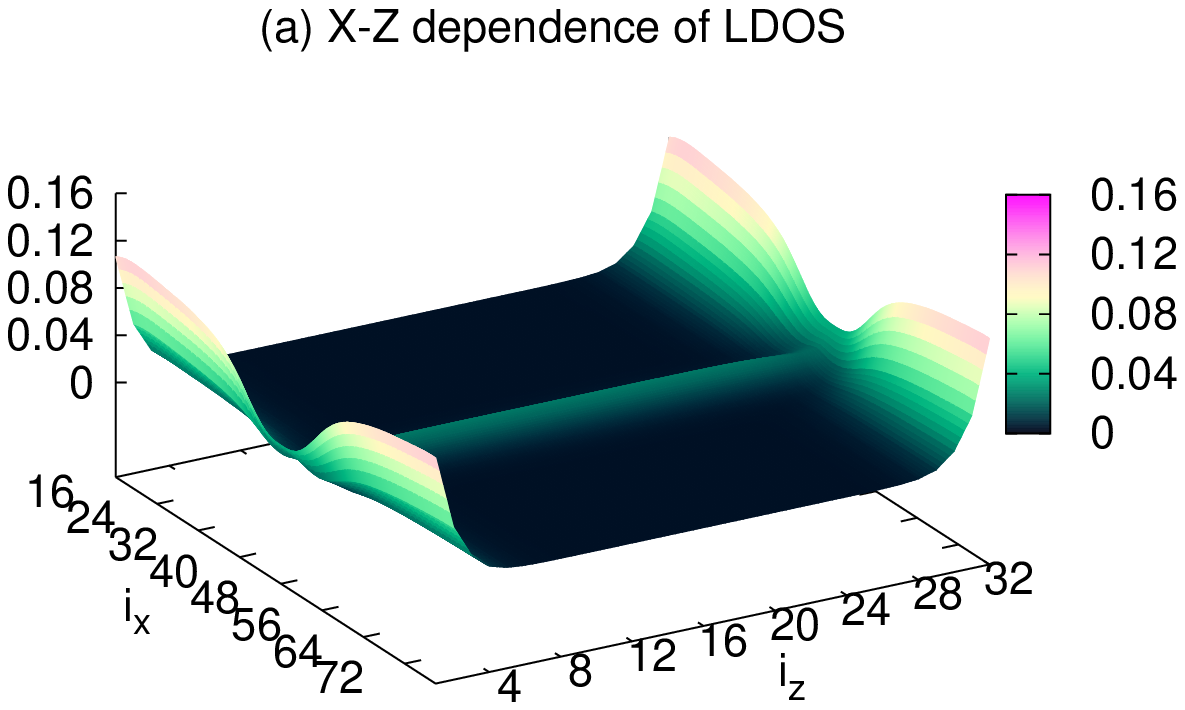}} 
      \begin{center}
      \resizebox{0.8 \columnwidth}{!}{\includegraphics{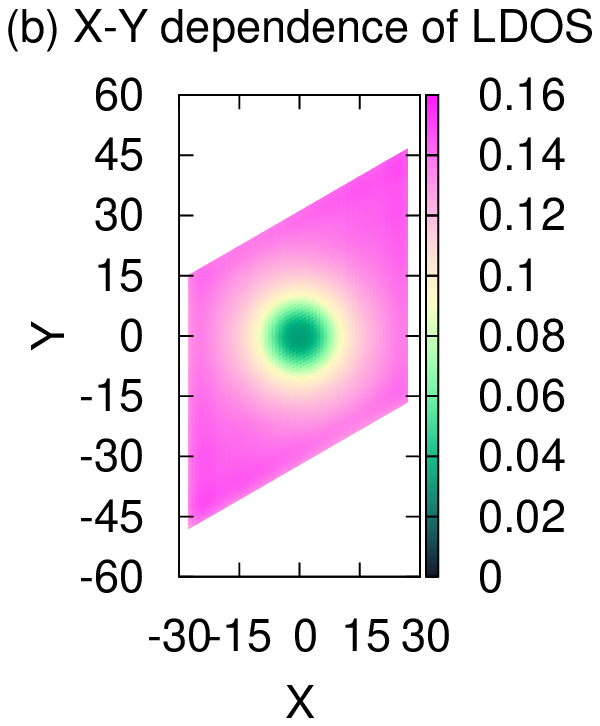}} 
      \end{center}
    \end{tabular}
\caption{\label{fig:fig4}
$X$-$Z$ and $X$-$Y$ dependence of the local density of states near zero energy $E/\Delta_{0} = -0.0067$. 
The vortex is located at the origin in $X$-$Y$ axes. 
}
  \end{center}
\end{figure}

Next, we focus on the vortex bound states. 
We calculate the LDOS on the line $i_{y} = L_{y}/2$ near zero energy ($E/\Delta_{0} = -0.0067$) as shown in Fig.~\ref{fig:fig4}(a). 
The vortex is located at $i_{x} = L_{x}/2 = 48$ in this system.  
The zero-energy bound states appear along the vortex line. 
We find that the intensity of the zero-energy vortex bound states are still finite even at the surface ($i_{z} = 1$ and $i_{z} = 32$). 
This numerical result suggests that the vortex bound states are robust against the surface. 
Although the vortex affects the surface bound states, 
the surface does not affect the vortex bound states at the intersection.
One would understand this behavior in terms of the symmetry which protects the vortex Majorana fermions. 
There is the mirror symmetry with respect to the plane perpendicular to the vortex\cite{Tsutsumi}. 
This mirror symmetry is not broken even with the surface\cite{Hsieh}. 

Finally, we focus on the LDOS on the surface near the zero energy. 
The LDOS around the zero energy is suppressed near the vortex as shown in Fig.~\ref{fig:fig4}(b). 
We note that the intensity of the LDOS at a vortex center is not zero. 
This finite intensity is originating from the vortex-bound states as shown in Fig.~\ref{fig:fig4}(a). 
Thus, the result suggests that the one-dimensional Majorana particle along a vortex is more robust than the two-dimensional Majorana particle on a surface. 
The suppression of the surface bound states can be observed by the scanning tunneling microscopy/spectroscopy (STM/STS). 
In conventional non-topological superconductors, the LDOS 
becomes large near a vortex, since the Andreev bound states are formed.
In topological superconductors, the surface bound states are suppressed by the presence of the vortex, 
which is similar to the behavior around a magnetic impurity on the surface of the topological insulator.
Our numerical calculation shows that the surface bound states are not robust against a vortex. 
In addition, we note that the down spin component of quasiparticles makes the bound states at the finite energy ($E/\Delta_{0} \sim -0.2$) as seen in Fig.~\ref{fig:fig3}(b) and (d), which means the existence of down-spin polarized vortex near a surface. 
This bound states can not be regarded as the vortex bound states shifted from the zero-energy level, since 
the vortex is up-spin polarized in the system without any surfaces as reported in our previous paper applying the magnetic field parallel to the $z$ direction\cite{NagaiMajo}. 
These down-spin polarized states can not be explained by the perturbation from the surface Majorana states, since 
the surface states are orbital-polarized as shown in Fig.~\ref{fig:fig3} so that these wave functions are different from that of the down-spin polarized states. 
The discussion based on a topology such as an index theorem can not also explain these states because these states are 
not zero-energy states, which means that the topological property does not protect these states.

In conclusion, we numerically  showed that 
the two-dimensional surface Majorana bound states are suppressed around a vortex perpendicular to the surface 
in an odd-parity fully-gapped topological superconductivity in massive Dirac BdG Hamiltonian, focusing on superconducting topological insulator Cu$_{x}$Bi$_{2}$Se$_{3}$ and Sn$_{1-x}$In$_{x}$Te. 
We also showed that vortex-bound states in topological superconductors are robust even with a surface.
The suppression of the surface bound states is similar to that with a magnetic impurity on the surface 
of the topological insulator.
The suppression of the surface-bound states around a vortex can be observed as the unconventional energy dependence of imaging of the scanning tunneling microscopy/spectroscopy.



\begin{acknowledgment}
We thank Y. Ota for helpful discussions and comments. 
The calculations have been performed using the supercomputing 
system PRIMERGY BX900 at the Japan Atomic Energy Agency. 
This study was partially supported by JSPS KAKENHI Grant Number 24340079 and 26800197. 
\end{acknowledgment}

\end{document}